\documentclass[aps,prl,reprint,superscriptaddress]{revtex4-1}
\usepackage{graphicx}
\usepackage{rotating}
\usepackage{amssymb}    
\usepackage{amsmath}   
\usepackage{epsfig}
\usepackage[normalem]{ulem}   
\usepackage{bm}   
\usepackage{color}

\date{\today}

\begin{document}

\title{The Boundaries of Synchronization in Oscillator Networks}

\author{Everton S. Medeiros}
\email{esm@if.usp.br}
\affiliation{Institute of Physics, University of S\~ao Paulo, Rua do Mat\~ao, Travessa R 187, 05508-090, S\~ao Paulo, Brazil}
\affiliation{Institute for Chemistry and Biology of the Marine Environment, Carl von Ossietzky University Oldenburg, Oldenburg, Germany}
\author{Rene O. Medrano-T}
\affiliation{Departamento de F\'isica, Universidade Federal de S\~ao Paulo,  Diadema, S\~ao Paulo, Brazil}
\author{Iber\^e L. Caldas}
\affiliation{Institute of Physics, University of S\~ao Paulo, Rua do Mat\~ao, Travessa R 187, 05508-090, S\~ao Paulo, Brazil}
\author{Ulrike Feudel}
\affiliation{Institute for Chemistry and Biology of the Marine Environment, Carl von Ossietzky University Oldenburg, Oldenburg, Germany}

\begin{abstract}
We analyze the final state sensitivity of nonlocal networks with respect to initial conditions of their units. By changing the initial conditions of a single network unit, we perturb an initially synchronized state. Depending on the perturbation strength, we observe the existence of two possible network long-term states: (i) The network neutralizes the perturbation effects and returns to its synchronized configuration. (ii) The perturbation leads the network to an alternative desynchronized state. By computing uncertainty exponents of a two-dimensional cross section of the state space, we find the existence of fractal basin boundaries separating synchronized solutions from desynchronized ones. We attribute these features to an unstable chaotic set in which trajectories persist for times indefinitely long in the network.
\end{abstract}
\maketitle

Synchronization is a universal concept in nonlinear dynamics characterizing the emergence of correlations between connected subsystems \cite{Kurths2003}. The ubiquity of synchronization transcends limits of different fields of science. In ecology, synchronization of populations in different habitats may be crucial for species persistence \cite{Stone1999}. In cardiology, the loss of synchronicity of the heart rhythms leads to the onset of ventricular fibrillation causing death \cite{Witkowski1998}. In neurodynamics, synchronicity of rhythms is a key concept to understand how processes performed in different parts of the neural system can be integrated \cite{Gray1994}. Also, many applications have been reported, e.g., mechanical oscillators \cite{Kapitaniak2014}, laser physics \cite{Thornburg1994}, phase oscillators \cite{Acebron2005} and complex networks \cite{Osipov2007,Arenas2008}. 

The asymptotic stability of synchronized states has been addressed in networks by the {\it master stability function} formalism \cite{Pecora1998,Pecora2000}. This method accounts for a linear analysis of perturbations applied transversely to the synchronized state. It provides satisfactory insights about the local stability of the synchronized state, however, it does not provide any information about its whole basin, i.e. the impact of larger perturbations beyond the linear limit. More recent approaches have focused on such global issues by estimating the relative size of the synchronization basin \cite{Girvan2006,Menck2013}. Although such analyses address a nonlocal character of stability in networks, yet, it does not give any insights about the boundaries of the basins of the synchronized state. The characteristics of such boundaries has practical consequences for the synchronization in networks. For instance, for fractal basin boundaries  \cite{Grebogi1985,McDonald1985}, the synchronized states would be sensitive to small perturbations in their initial conditions, even when these states are classified as asymptotically stable by the previously mentioned formalisms. Surprisingly, final state sensitivity due to fractal basin boundaries has not been addressed for synchronized states so far.

We investigate the sensitivity of synchronized states with respect to initial conditions in homogeneous networks and show that a small perturbation in the initial condition of a single unit is capable to desynchronize the whole network. The sole requirement for each network unit is the coexistence of an unstable chaotic set with a periodic attractor. As consequence, two long-term network states are available, a complete synchronized state at the periodic attractor and a fully desynchronized one around the chaotic set which appears to turn into an attractor. We explain this behavior by employing the concept of final state sensitivity. The final state sensitivity appears in two regimes: A less uncertain regime corresponding to the boundaries of a continuous region of perturbations leading to synchronization; and an extremely sensitive regime corresponding to an intricated structure of perturbations. The uncertainty exponents of both regimes indicates the occurrence of fractal basin boundaries between synchronized and desynchronized states. We point out the generality of this behavior, since coexistence of unique attractors and chaotic saddles embedded in their basins of attraction is a generic feature of nonlinear oscillators. 

We study a network composed of $N$ units arranged in a ring. For the dynamics in each unit we chose the Duffing oscillator: 
\begin{eqnarray}
      \label{Eq1}
    \dot{x}_{i}  &=& y_{i} + \frac{\sigma}{2R}\sum_{j=i-R}^{j=i+R}(x_{j}-x_{i}), \\
      \dot{y}_{i}  &=& -\gamma y_{i} + x_{i} - x_{i}^{3}+A \cos(\omega t) +\frac{\sigma}{2R}\sum_{j=i-R}^{j=i+R}(y_{j}-y_{i}).
         \nonumber
          \end{eqnarray}
The pair $(x_{i},y_{i})$ defines the state space of each oscillator $i$ with $i=1,...,N$.  The parameter $\sigma$ controls the coupling strength, $R>1$ specifies the number of first neighbors connected to the $i$th unit, realizing a nonlocal coupling of the units. The parameter $\gamma$ denotes the damping in the system, while $A$ and $\omega$ are the forcing amplitude and frequency, respectively. We fix the parameters of each unit at $\omega=0.5$, $\gamma=0.24$, and $A=13.633$, where the Duffing oscillator exhibits a globally stable period-$3$ attractor, ${\bf A}$, and an unstable chaotic set, $\Lambda$. The states $(x_{i},y_{i})$ are sampled at multiples of the forcing period, $T=2\pi/\omega$. The coupling parameters $R$ and $\sigma$ as well as the network size $N$ are specified for each simulation.    

\begin{figure}[!htp] 
\centering
\includegraphics[width=8.5cm,height=4cm]{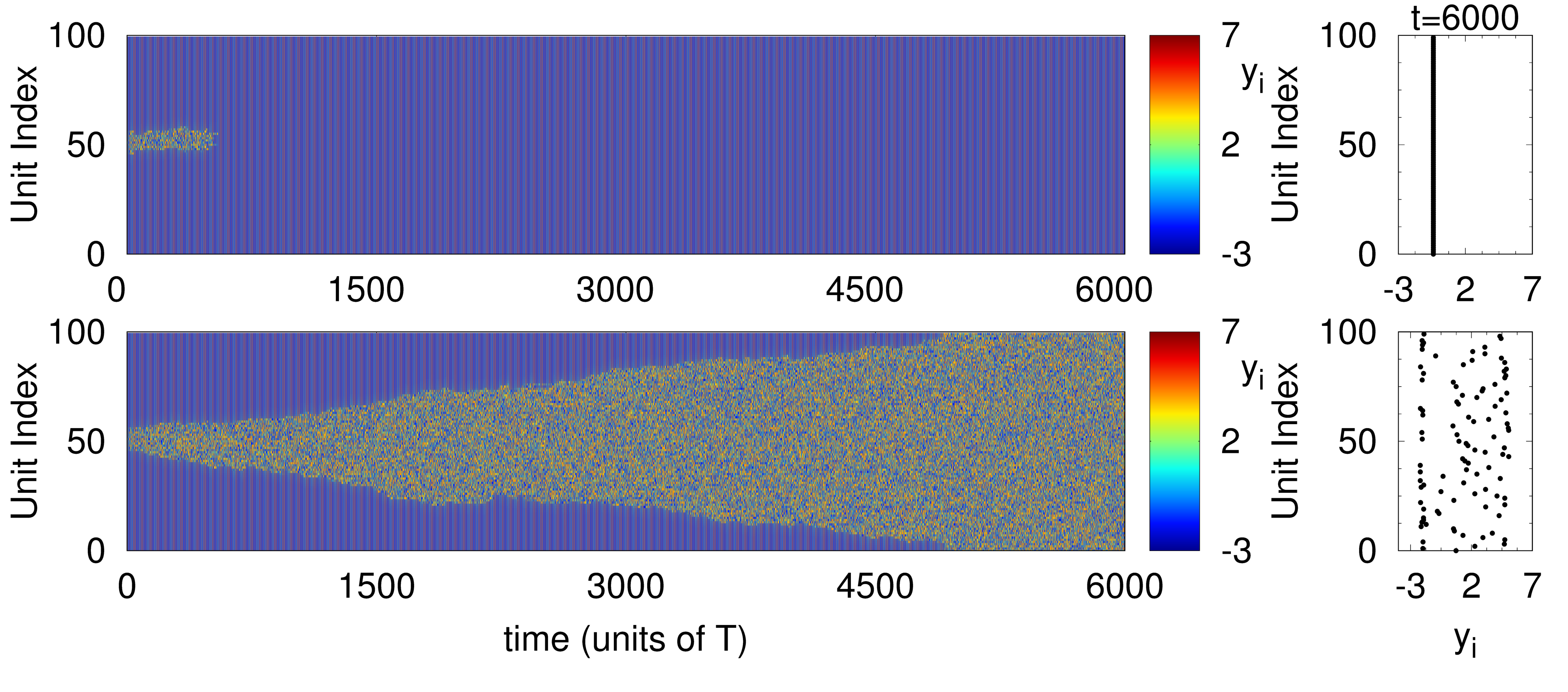} 
\caption{Space-time plots and time snapshot of system~(\ref{Eq1}). (top) Initial configuration for which the network reaches full synchronization following a period-$3$ orbit. (bottom) Initial configuration for which the network completely desynchronizes. The coupling parameters are $R=5$ and $\sigma=0.004$, the network size is $N=100$.}
\label{figure_1}
\end{figure}

In Fig.~\ref{figure_1}, we show the dynamics of the network in space and time described by system~(\ref{Eq1}). In these simulations, all network units, but one, are set to the same initial condition defining a synchronization manifold, ${\bf S}$. All units that compose the synchronization manifold started at ($-5.0$, $5.0$). In Fig.~\ref{figure_1}(top), the unique unit starting outside the synchronization manifold receives ($-2.25$, $-11.00$) as initial condition. This setup leads to synchronization of the whole network after some transient time, as illustrated with snapshots ($i \times y_i$) in the right. Next, the initial condition of the unique unit is slightly perturbed to ($-2.25$, $-11.00+\epsilon$) with $\epsilon=10^{-4}$. As demonstrated in Fig.~\ref{figure_1}(bottom), the network converge to a completely desynchronized state, see snapshot in the right. Hence, Fig.~\ref{figure_1} demonstrated the coexistence of two different global network behaviors, a fully synchronized and a completely incoherent one. Moreover, it suggests that the network's final state is highly sensitive with respect to small changes in its initial conditions. 

The synchronized solution shown in Fig.~\ref{figure_1}(top) possesses a basin of synchronization, i.e., a set of initial conditions that leads the network to the synchronized state. We compute a slice of such basin by keeping all initial conditions of ($N-1$) constant at the chosen synchronization manifold and vary only the initial conditions attributed to the perturbed unit. In order to identify such synchronization basin, we define an order parameter $\mathcal{Z}$ by analyzing a next neighbor error $E_i=\|\bm{r}_i-\bm{r}_{i-1}\|$, where $\bm{r}_i$ is the state vector of the dynamical variables $(x_i,y_i)$;
\begin{eqnarray}
\mathcal{Z}=\frac{1}{N}\sum_{i=1}^{N}K_i, \;\;\; K_i=\begin{cases} 1,& \text{if} \;\; E_i > \delta\\
    0,              & \text{if} \;\; E_i < \delta.  \end{cases}                                                                                           
\end{eqnarray}
The parameter $\delta$ establishes the synchronization quality, which we fixed to $\delta=0.01$. Hence, the completely synchronized state returns an order parameter $\mathcal{Z}=0$, while the completely desynchronized state implies an order parameter $\mathcal{Z}=1$.

\begin{figure}[!htp] 
\centering
\includegraphics[width=8cm,height=6cm]{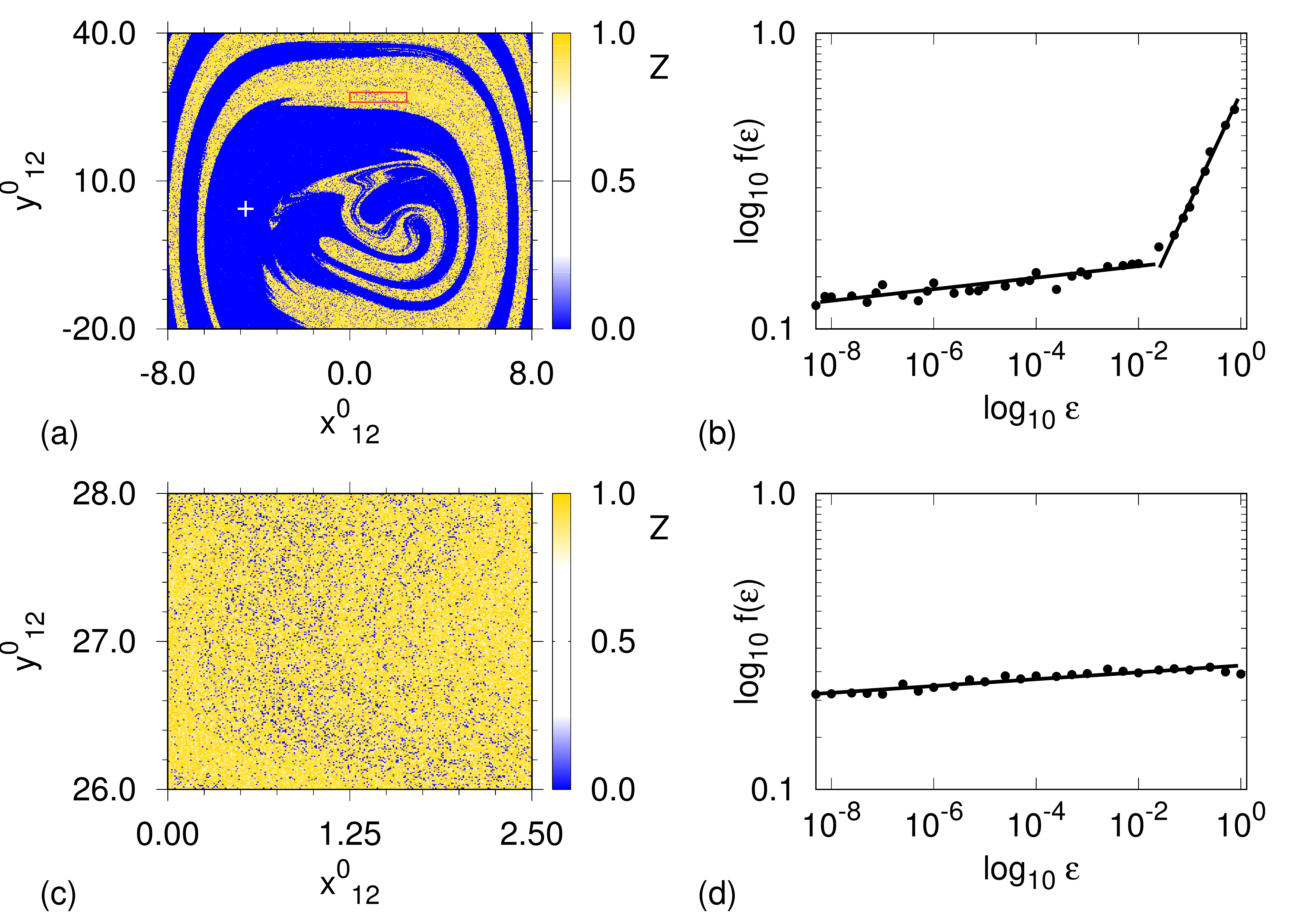}
\caption{($a$) Domain of initial conditions attributed to the unique element outside the synchronization manifold ($x_{12}^{0}$,$y_{12}^{0}$). The values codified in blue (dark gray) correspond to unit initial conditions leading to synchronization, while initial conditions in yellow (light gray) correspond to desynchronized states. The white cross indicates the initial conditions of rest $(N-1)$ units defining the synchronized state. Network parameters are $N=25$ and $\sigma=0.004$. ($b$) The fraction $f(\varepsilon)$ of initial conditions that are uncertain under the perturbation $\varepsilon$. ($c$) Magnification of the red rectangle of ($a$). ($d$) The corresponding  $f(\varepsilon)$ of ($c$).}
\label{figure_2}
\end{figure}

The order parameter $\mathcal{Z}$ is computed for a $2$-dimensional grid of initial conditions ($x_{12}^{0}$, $y_{12}^{0}$) attributed to the  single perturbed unit of a network composed of $N=25$ units. Accordingly, the plane spanned by $x_{12}^{0}$ and $y_{12}^{0}$ forms a two-dimensional section of the $50$-dimensional state space. For each initial condition on this plane, a transient initial time of $2 \times 10^{4}T$ is discarded before computing the order parameter shown in Fig.~\ref{figure_2}($a$). The initial conditions colored in blue (dark gray) correspond to the basin of the synchronized state, while the ones colored in yellow (light gray) correspond to trajectories for which the whole network desynchronizes. The white cross indicates the initial condition attributed to the synchronization manifold, i.e., the ($N-1$) network units defined as synchronized in the initial instant of time. Since we have fixed the overall integration time to $t_{end}= 2 \times 10^{4}T$, some order parameters, white in Figs.~\ref{figure_2}($a$) and ($c$), can take values between $0$ and $1$ this corresponds to network realizations for which full (de)synchronization has not been reached yet. Those are initial conditions corresponding to unstable chimera states \cite{Kuramoto2002,Strogatz2004,Omelchenko2011}.
 
As we observe in Fig.~\ref{figure_2}($a$), the basin of the synchronized state contains open sets of initial conditions in which all of them converge to synchronization in the period-$3$ orbit; they appear as blue (dark gray) continuous areas. However, we note that these open sets are separated by a nontrivial boundary from complex regions, where a mixture of blue (dark gray) and yellow (light gray) points occurs indicating high final state sensitivity. In those mixed regions, Fig.~\ref{figure_2}($c$), small changes in the initial condition of the perturbed unit lead to the incoherent state of the whole network. This part of the basin corresponds to a cross section of the extended state space in which the stable manifold (foliation) of the chaotic set is close.

We computed the fraction of initial conditions that are uncertain with respect to perturbations $\varepsilon$ \cite{Grebogi1985,McDonald1985}. An initial condition is called uncertain if it converges to some specific attractor when unperturbed, but converges to another when a perturbation of size $\varepsilon$ is applied. Considering a large interval for such perturbations, $\varepsilon \in [10^{-8},10^{0}]$, the fraction of uncertain initial conditions $f(\varepsilon)$ scales with $\varepsilon$ as $f(\varepsilon) \sim \varepsilon^{\alpha}$, where $\alpha$ is the uncertainty exponent. Fig.~\ref{figure_2}($b$) shows the relation $\log_{10}(f(\varepsilon))$ versus $\log_{10}(\varepsilon)$ for the domain of attraction shown in Fig.~\ref{figure_2}($a$). We find two different uncertainty regimes in the interval of $\varepsilon$. For small values of $\varepsilon$, including the limit $\varepsilon \rightarrow 0$, the uncertainty exponent is significantly lower than the one for larger perturbations. The lower exponent, $\alpha=0.019 \pm 0.002$, indicates extreme sensitivity to perturbations expected for the complex region, while the larger exponent, $\alpha=0.370 \pm 0.002$, indicates lower uncertainty of the open sets. This scale separation becomes more clear if we restrict the region in state space for which we compute the uncertainty exponent, shown in Fig.~\ref{figure_2}($c$) (the rectangle in Fig.~\ref{figure_2}($a$)), yielding $\alpha=0.011 \pm 0.002$, (Fig.~\ref{figure_2}($d$)). The proximity of the uncertainty exponent to zero, indicates the occurrence of riddled basins \cite{Alexander1992}. However, the strict definition of riddled basins is valid for the coexistence of at least two attractors, where the probability of reaching one of them is positive within any neighborhood of the other. Here, due to the difficulty of strictly ascertain the desynchronized state as an attractor of the system, we classify the proximity of the uncertainty exponent to zero, $\alpha \approx 0$, as an indication of riddled-like basins \cite{Woltering2000}.

\begin{figure}[!htp] 
\centering
\includegraphics[width=8.9cm,height=4cm]{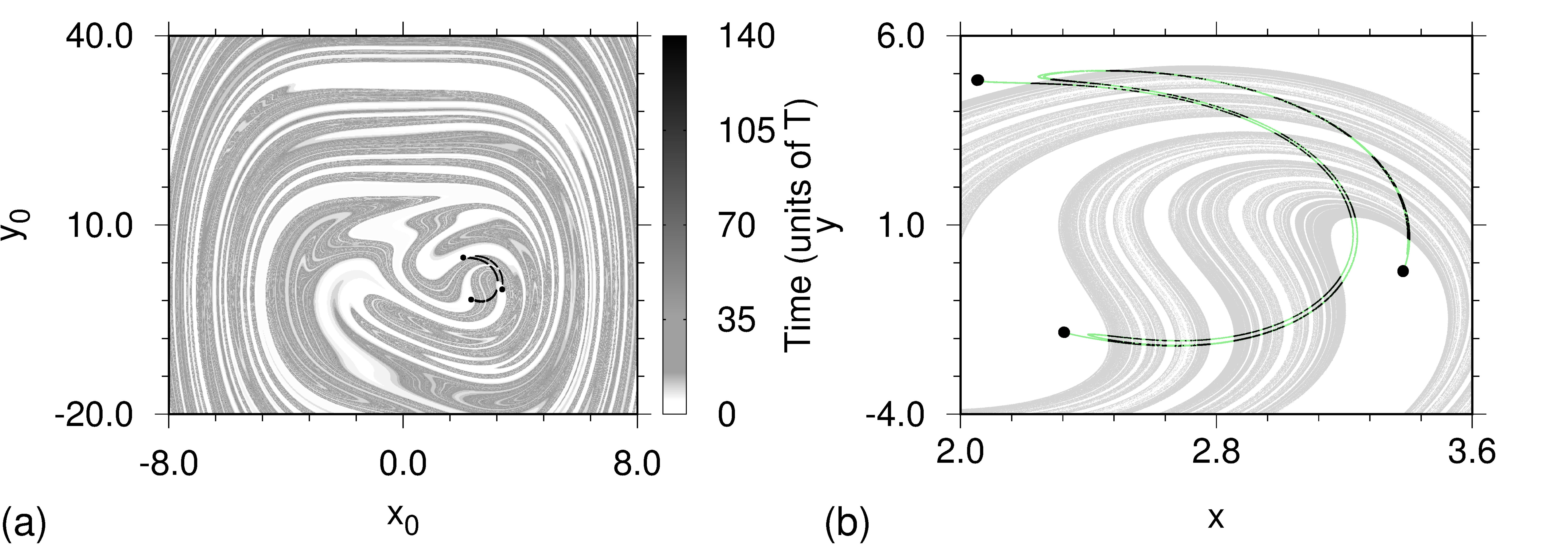}
\caption{(a) Domain of attraction of the unique period-$3$ attractor of the Duffing oscillator. The attractor is represented by the black circles. The the shades of gray codifies the time spent for trajectories to reach the attractor, darker shades corresponds to trajectories crossing the chaotic saddle which is indicated in black. (b) Stable (gray) and unstable manifold (green (light gray)) of the chaotic saddle (black).}
\label{figure_3}
\end{figure}

The mechanism leading to the two distinct behaviors of the network lies in the existence of an unstable chaotic set $\Lambda$, a chaotic saddle, which is embedded in the state space of each unit of the uncoupled system and also persists in the high dimensional state space of the network. As pointed out previously, each unit possesses a globally stable attractor $\textendash$ a period-$3$ orbit (black circles in Figs.~\ref{figure_3}($a$) and ~\ref{figure_3}($b$)) $\textendash$ besides the chaotic saddle (black dots in Fig.~\ref{figure_3}($a$) and Fig.~\ref{figure_3}($b$)). Though all initial conditions ultimately converge to the period-$3$ orbit, the time they need to reach the neighborhood of the attractor can be quite different, as demonstrated by the shades of gray according to the length of their transient time. We find that those long convergence points (colored black) belong to a set very close to the stable manifold of the chaotic saddle embedded in the basin of attraction. Using the sprinkler method \cite{Tamas2011}, we obtain an approximation of that stable manifolds, gray dots in Fig.~\ref{figure_3}($b$). All trajectories starting from initial conditions very close to it will first approach the chaotic saddle, dwell close to it for some time interval and finally will be ejected along its unstable manifold (green (light gray) in Fig.~\ref{figure_3}($b$)) reaching the period-$3$ attractor. This dynamical feature of the state space of a single unit has striking consequences for the network dynamics.

\begin{figure}[!htp] 
\centering
\includegraphics[width=8cm,height=3.4cm]{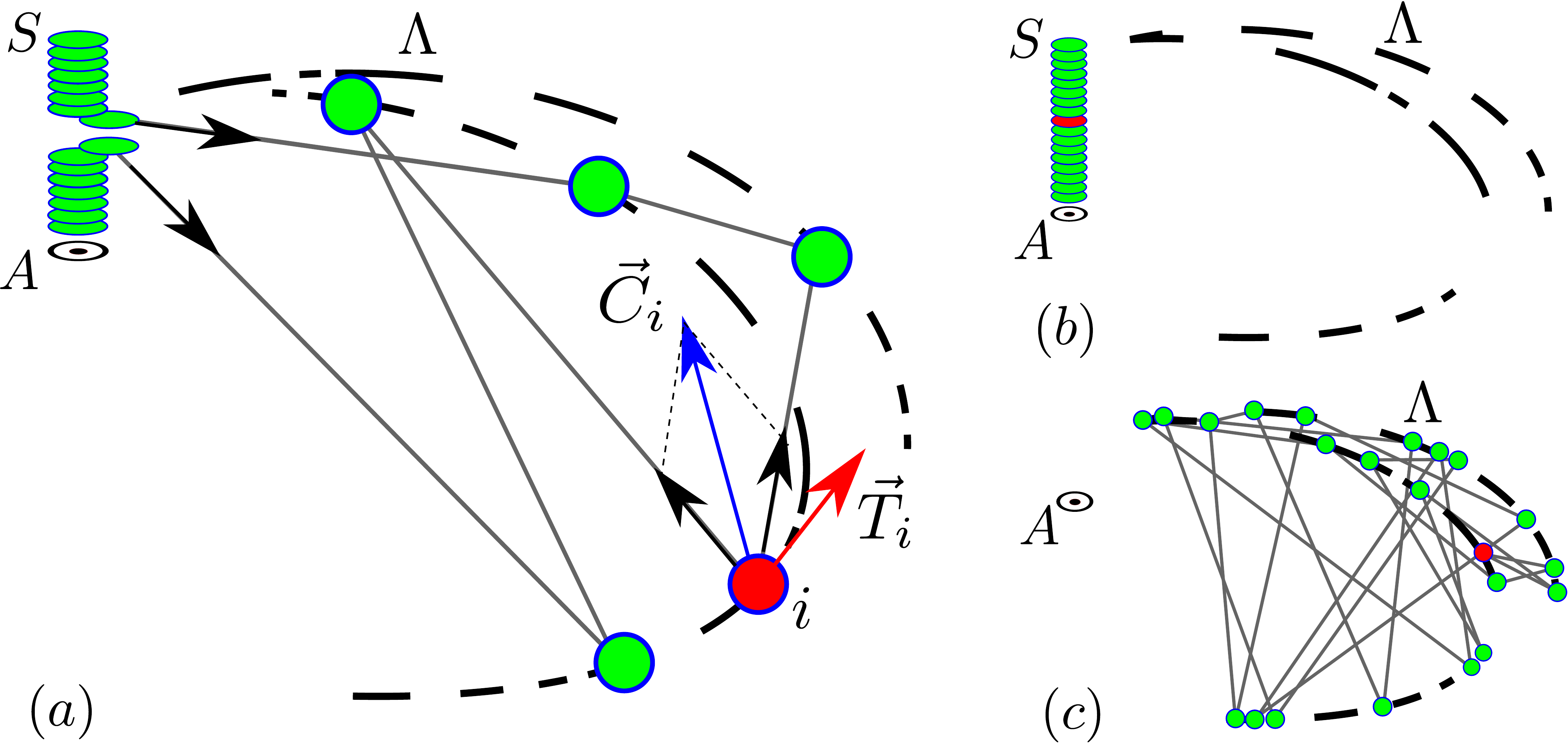} 
\caption{($a$) Schematic of a cluster of units in the chaotic set, $\Lambda$. The red (dark gray) circle represents the perturbed unit. The arrow $\vec{T}_i$ indicate the local dynamics of the unit $i$ in the chaotic set, while the arrow $\vec{C}_i$ indicate the resulting coupling on the unit $i$. ($b$) Completely synchronized final state. ($c$) Fully desynchronized long-term state.}
\label{figure_4}
\end{figure}

All network units starting with synchronized initial conditions that approach the chaotic set, i.e, they are close to its stable manifold, would first approach the chaotic set in a synchronized manner and then converge to the period-$3$ orbit through its unstable manifold. For such features, if the unique perturbed unit approaches the chaotic saddle, its originally synchronized neighbors will be distorted due to exponential separation of nearby trajectories on the chaotic saddle leading to their desynchronization. This mechanism is shown in the sketch of Fig.~\ref{figure_4}($a$), where a cluster of units is close to the set $\Lambda$, and the other units have reached the attractor ${\bf A}$. Every network unit in the cluster is subject to two different influences: their local dynamics represented by $\vec{T}_i$ (see arrows in Fig.~\ref{figure_4}($a$)); and the resulting coupling, $\vec{C}_i$, due to the coupling to $R$ neighbors on each side (arrow in Fig.~\ref{figure_4}($a$)). Since the overall coupling on any unit depends on the instantaneous trajectories of its neighbors, the trajectory of every unit in the cluster is constantly influenced by the motion on the chaotic set. Such sensitive disturbances may retard the escape of units from the chaotic saddle especially by decreasing the likelihood of approaching its unstable manifold \cite{Kantz1985,Lai1995}. Due to the coupling with units outside the initial cluster, even more units can be pulled into the chaotic saddle increasing the cluster of desynchronized units. Once a critical number of desynchronized units is reached, the convergence to full desynchronization becomes possible (Fig.~\ref{figure_4}($c$)) \cite{Note1}. Otherwise, all network units end up completely synchronized at the periodic attractor, Fig.~\ref{figure_4}($b$). 

\begin{figure}[!htp] 
\centering
\includegraphics[width=8.9cm,height=4cm]{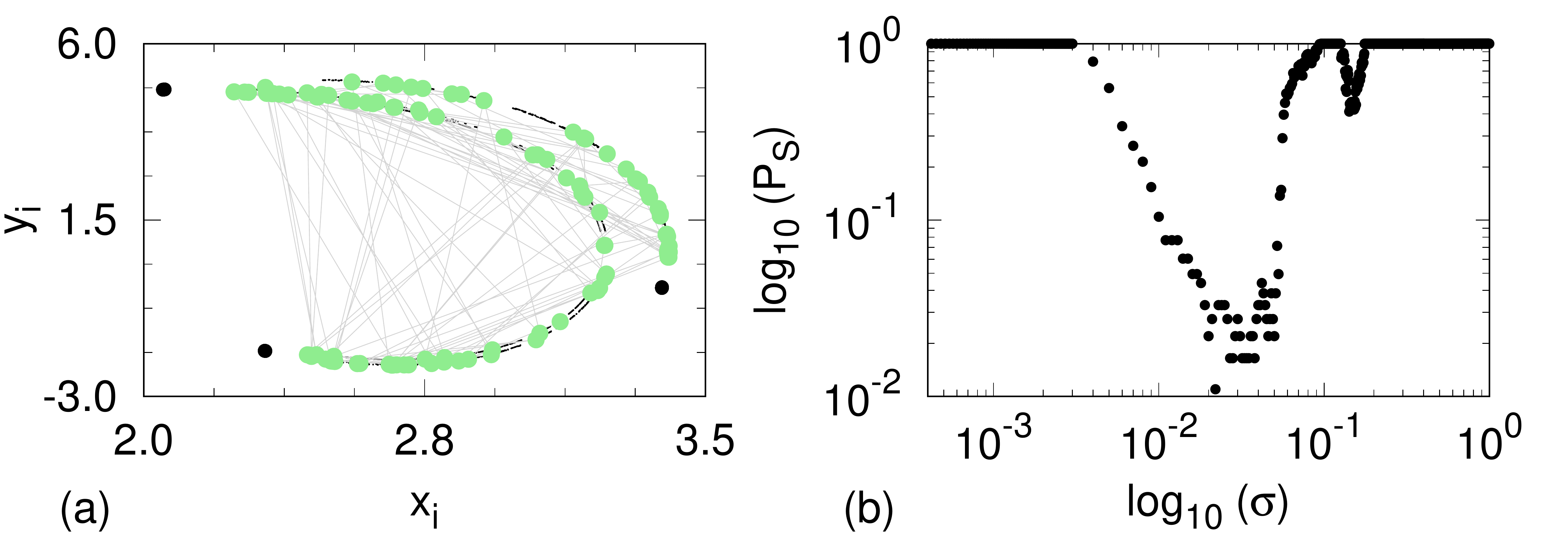} 
\caption{($a$) In black is the chaotic set, $\Lambda$, of the Duffing oscillator. The green (gray) circles corresponds to a snapshot of the pair ($x_i,y_i$) of each unit at time instant $t=10^6 T$. Black circles illustrate the period-$3$ attractor. Network parameters are $N=100$, $R=5$ and $\sigma=0.004$ ($b$) Probability of completely synchronized states as function of the coupling strength $\sigma$. $N=25$ and $R=5$.}
\label{figure_5}
\end{figure}

Now, we investigate the structure of the fully incoherent state. Fig.~\ref{figure_5}($a$) shows the state variables of each unit after a very long time of network iterations, $10^6$ cycles of the forcing $T$. We observe that in the fully desynchronized state, each network unit is tracing out the chaotic set, indicating its dominance in the dynamics. This result suggests that the chaotic set, unstable for a single unit, plays the role of an attractor in the network. 

Finally, we address the question how robust this phenomenon is with respect to the variation of the coupling strength. To this end, we vary $\sigma$ in the interval [$0$,$1$], and perform network simulations with different initial conditions attributed to the perturbed unit. We compute the probability of convergence to the synchronized state, $P_S=n_S/n$, as a function of $\sigma$ for $n=182$ realizations discarding a large interval of transients (Fig.~\ref{figure_5}($b$)). Here, $n_S$ is the number of realizations that leads to network synchronization. We remark that in the first interval, $\sigma$ $\in$ [$0$,$0.0036$], for very small values of coupling constants $\sigma$, network units may reach the attractor with different phases. However, even for intermediate values of $\sigma$ in this interval, complete synchronization is already observed. As we increase further the coupling constant, we find an approximated interval,  $\sigma$ $\in$ [$0.0036$,$0.18$], in which the probability of observing the completely synchronized state is lower than $1$. This interval corresponds to the coupling regime for which the loss of synchrony leading to an incoherent state is observed. For values of $\sigma$ even larger, the coupling easily overcomes the chaotic dynamics in the neighborhood of the chaotic set and the network completely synchronizes for all network realizations, $P_S=1.0$. 

\begin{figure}[!htp] 
\centering
\includegraphics[width=5cm,height=3.5cm]{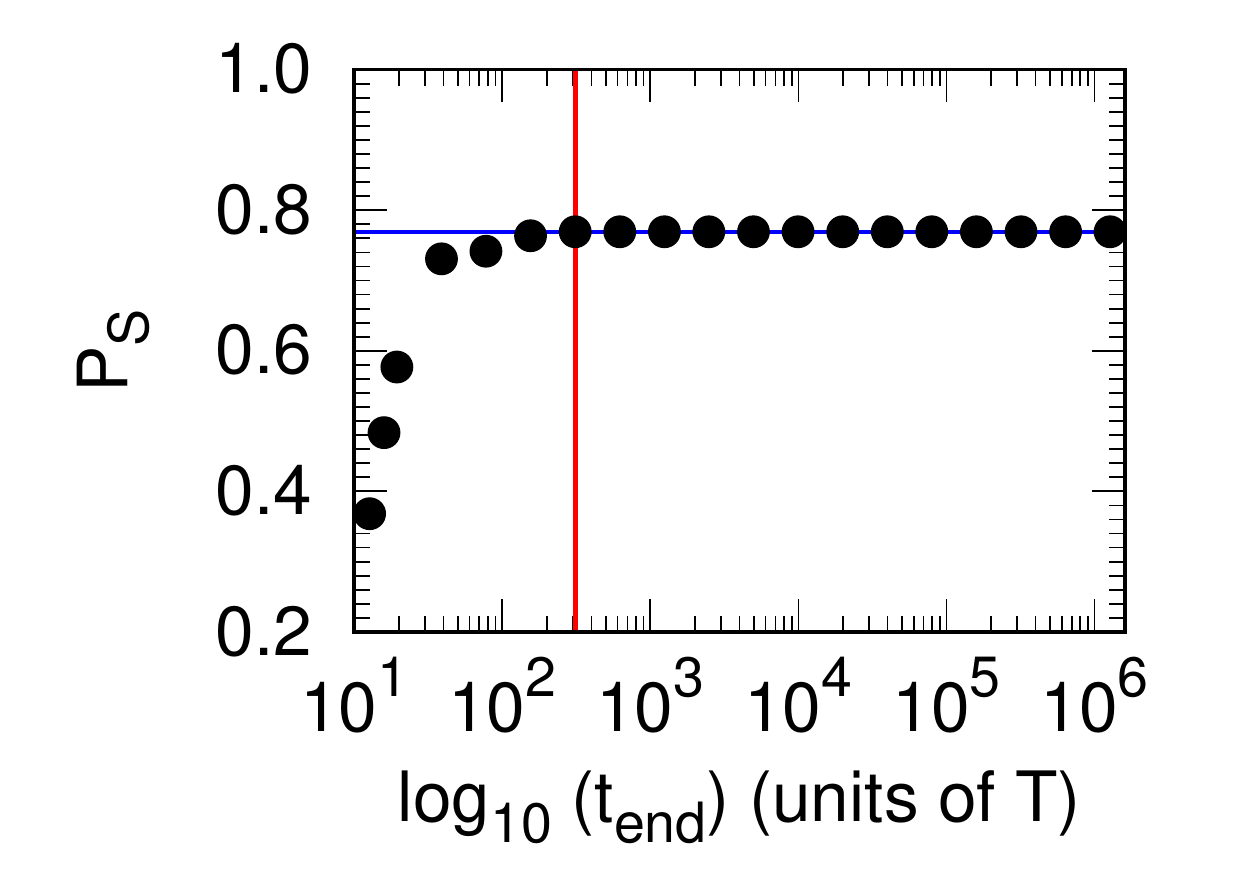} 
\caption{The probability of complete synchronization as function of different simulation times, $t_{end}$, in units of the period of the system forcing, $T$. The blue (horizontal) line indicates the constant probability, $P_S\approx0.77$. The red (vertical) line indicate the time, $t_{end}\gtrsim 311$, for the probability $P_S$ to stabilize. Network parameters are $N=25$, $R=5$, and $\sigma=0.004$.}
\label{figure_6}
\end{figure}

To check the persistence of the incoherent behavior, and to distinguish it from chaotic transients, we again consider $n=182$ simulations, each one starting with different, randomly chosen, initial conditions attributed to the perturbed unit. By computing the number of networks, $n_S$, for which the network synchronizes as a function of different simulation times, $t_{end}$, we obtain the probability of complete synchronization, $P_S=n_S/n$. Fig.~\ref{figure_6} shows that $P_S$ saturates at a fixed value, $P_{S}\approx0.77$, while the complementary probability of $P_S$ corresponding to the likelihood of desynchronization, reaches $P_{\Lambda}=1-P_S\approx0.23$. For conventional transient dynamics, the probability $P_S$ is expected to approach $1$ as the simulation time, $t_{end}$, increases. In our case, the stability of $P_{S}\approx0.77$ for $t_{end}\gtrsim 311$ $T$, indicates that incoherent state is present for arbitrarily long times. Hence, in this case, it is impossible to ascertain this behavior as a transient as done for spatially extended system in Refs. \cite{Tamas2011,Lilienkamp2017,Lilienkamp2018}. Here, the chaotic saddle of the single unit appears to turn into an attractor. Numerically, this is difficult to distinguish from a super-persistent chaotic saddle which would be characterized by a large unlikeliness of unstable directions \cite{Kantz1985,Crutchfield1988,Lai1995}. 

We reported the existence of final state sensitivity between synchronized and desynchronized states in networks with nonlocal coupling. It indicates the occurrence of fractal basin boundaries between these two solutions. The observed network sensitivity appears in two regimes: a moderate sensitivity corresponding to fractal boundaries of a continuous region of initial conditions leading to synchronization; and a very sensitive regime corresponding to riddled-like region of initial conditions leading to synchronized and desynchronized states. This is due to the interplay of the coupling and the unstable chaotic sets. Furthermore, we observe that the desynchronized state appears to be an attractor or a super-persistent chaotic saddle with a low likelihood of unstable directions. Finally, we emphasize that these findings are independent of a particular choice of the dynamics for the network units, since the coexistence between an unstable chaotic set and a periodic attractor is sufficient for the occurrence of the results reported here. Such a situation is generic for almost all chaotic systems, since it is present in any periodic window interspersed into chaotic dynamics. The same behavior can be expected for excitable systems possessing period-adding cascades in which parameter ranges of mixed mode oscillations are interspersed with chaotic regions. Examples include the spike adding dynamics in bursting neuron models like the Hindmarsh-Rose \cite{Barrio2014} and a Hodgkin-Huxley model for thermally sensitive neurons \cite{Feudel2000}.

The authors thank Professor T. T\'el for useful discussions. This work was supported by FAPESP (Process: 2011/19296-1, 2013/26598-0, 2015/50122-0 and 2017/05521-0 ).

\end{document}